\documentstyle[aps,twocolumn,epsf]{revtex}

\begin{document}
\draft

\title{ Inter- and Intra-Chain Attractions in Solutions of
Flexible Polyelectrolytes at Nonzero Concentration }

\author{J.C. Chu and C.H. Mak}
\address{ Department of Chemistry, University of Southern California,
  Los Angeles, California 90089-0482, USA }

\date{Date: \today}
\maketitle


\begin{abstract} 

Constant temperature molecular dynamics simulations were used to study
solutions of flexible polyelectrolyte chains at nonzero concentrations 
with explicit counterions and unscreened coulombic interactions.  
Counterion condensation, measured via the self-diffusion 
coefficient of the counterions, is found to increase 
with polymer concentration, 
but contrary to the prediction of Manning theory, 
the renormalized charge fraction on the chains decreases 
with increasing Bjerrum length without showing any saturation.
Scaling analysis of the radius of gyration shows that the chains are 
extended at low polymer concentrations and small Bjerrum lengths, 
while at sufficiently large Bjerrum lengths, the chains shrink to produce 
compact structures with exponents smaller than a gaussian chain, 
suggesting the presence of attractive intrachain interactions.
A careful study of the radial distribution function 
of the center-of-mass of the polyelectrolyte chains shows clear
evidence that effective  interchain attractive interactions
also exist in solutions of flexible polyelectrolytes, similar
to what has been found for rodlike polyelectrolytes.
Our results suggest that the broad maximum observed in scattering 
experiments is due to clustering of chains.

\end{abstract}

\narrowtext

\section{Introduction}\label{sect1}

Polyelectrolytes are long-chain molecules with ionizable side-groups
\cite{oozawa,barrat-joanny}.  
In a polar solvent, these ionizable groups can dissociate resulting 
in localized charges along the backbone of the polymer forming a ``macroion'', 
plus mobile counterions that are no longer permanently
bonded to the chain.  Due to the long-range nature of the 
electrostatic interactions, the counterions, though mobile, are not entirely 
uncorrelated with the macroions.  For polyelectrolytes with a sufficiently 
high charge density along the backbone, some of the counterions remain 
bound to the chain.  This phenomenon is known as ``counterion condensation''.

The qualitative theory of counterion condensation has been worked out by 
Manning \cite{manning} for long rodlike polyelectrolytes at infinite 
dilution.  Within the Manning theory, counterion condensation occurs when 
the distance between charges $b$ along the chain is sufficiently small
(i.e. the charge density is sufficiently high) compared to the length scale
set by the electrostatic interactions $\lambda_B = e^2/\varepsilon k_B T$ ($e$ being 
the charge, $\varepsilon$ the dielectric constant and $k_B T$ the thermal 
energy).  Counterions continue to condense
onto the macroions, resulting in a renormalized charge density along each 
chain, until the effective average charge separation $b_{\rm eff}$ equals 
$\lambda_B$.

Manning's theory provides a lucid picture for understanding counterion
condensation, but its simplicity invites more complex questions.  For 
example, it is not clear how the qualitative features of the theory
would be modified by having flexible chains.  Equally unclear is 
how the essential results of the theory can be generalized to nonzero 
concentrations instead of just at infinite dilution.  
Recent experiments \cite{amis,ermi,williams}, theories 
\cite{mandel,muthukumar,liu1,becky} and simulations 
\cite{stevens-kremer,winkler} have
attempted to address some of these issues, but no consistent picture
has yet emerge.

Another issue related to counterion condensation concerns the possibility
of attractive interactions between two or more like-charged chains.  
There is now convincing evidence, both numerical \cite{gelbart} and 
analytical \cite{liu2}, that attractive interactions can exist between 
rodlike polyelectrolytes with the same charge.  This attraction is 
mediated by the fluctuating charge density along the rods created by the 
condensed counterions when they move on and off the rods, leading to 
an effective van der Waals-like interaction.  But the same question for
flexible polyelectrolyte chains is still open.  Since flexible chains 
can undergo conformational changes as the degree of counterion condensation
changes and deviation from linearity in turn affects the degree of
counterion condensation, these two factors become highly convoluted and their
net effects on the attractive interaction is unclear.  

This paper will try to address some of the issues raised here for 
solutions of flexible polyelectrolytes at nonzero concentrations.  The
complexity of the problem demands a careful treatment of the relevant 
microscopic interactions involved.  Computer simulations provide a 
simple way to study the problem without resorting to approximate
methods such as the Debye-H\"uckel or other mean field theories.  
In the simulations described below, 
all counterions in the solution as well as charges 
on the macroions are handled explicitly, with the full unscreened coulombic
interactions among them.  The method we used is similar to those in 
other computer simulations reported recently \cite{stevens-kremer}.  
While the focus of the previous studies has been on chain structure and
single-chain properties, the present study concentrates on inter- 
and intra-chain interactions.

\section{Model and Simulation Method}\label{sect2}

In our simulations, each polyelectrolyte chain is represented by a bead-spring
model with $N$ monomers.  Every monomer has unit charge $Z=1$.  We employ
cubic periodic boundary condition, with $M$ polymer chains inside the 
simulation box of dimension $L^3$.
In this study, we have considered only solutions with 
monovalent counterions of charge $Z=-1$ and no added salt; 
therefore, the number of counterions $N_c$ necessarily equals $M \times N$,
due to charge neutrality.  
No explicit solvent molecules are included in the simulations.
The solvent is represented by a dielectric continuum.  
The particles interact with each other through a coulombic potential, 
renormalized by the dielectric constant $\varepsilon$.
In addition to the coulombic interaction, every pair of particles has  
an excluded volume interaction, represented by a 
purely repulsive truncated and shifted Lennard-Jones (RLJ) potential
\begin{equation} \label{RLJ}
 U^{\rm RLJ}(r) = \left\{ \begin{array}{ll}
                   4\epsilon[(\frac{\sigma}{r})^{12} -
                               (\frac{\sigma}{r})^6  +
                                \frac{1}{4}], 
                              & \mbox{ $r \leq 2^{1/6} \sigma$ }; \\
                     0, & \mbox{ $r > 2^{1/6} \sigma$ }.
                     \end{array}
             \right.
\end{equation}
The monomer size is approximately $\sigma$.  We measure all lengths in
units of $\sigma$ and all energies in units of $\epsilon$.  
The connectivity of the chains is maintained by the
finitely extensible nonlinear elastic (FENE) potential \cite{FENE}:
\begin{equation} \label{FENE}
U^{\rm FENE}( r) = \frac{1}{2} k R_0^2 \ln (1-r^2/R_0^2),
\end{equation}
which defines the spring potential between consecutive monomers along the
chain.  We set $k=7\epsilon/\sigma^2$ and $R_0 = 2\sigma$ to ensure 
essentially no bond crossing.  We assume the polyelectrolyte backbone is
in a good solvent, therefore no explicit solvent-polymer interactions has
been included.  This model as well as the parameters are essentially
identical to those used in previous simulations by Stevens and Kremer
\cite{stevens-kremer}.

Special handling of the coulombic interaction is necessary in 
a solution with finite concentration.  The minimum image convention is usually
sufficient for short range interactions, but the long range nature
of the coulomb potential causes a pair of charges to interact far beyond 
their first periodic image.  The Ewald summation method properly accounts for
such contributions from all the images, but 
its implementation generally requires extensive 
computation.  An efficient alternative is the tabulated Ewald 
method with interpolation \cite{tab-Ewald} 
which we have used in the present study.  
This allows us to treat arbitrarily high concentrations 
with high efficiency and 
without approximation.

The scale of the charge-charge interaction is set by the Bjerrum length 
$\lambda_B=e^2/\varepsilon k_B T$.  
The simulations have all been carried out at
a constant temperature $k_BT= 1.2 \epsilon$.  The Bjerrum length can 
be varied by changing the dielectric constant $\varepsilon$.  For the 
results reported below, $\lambda_B$ varies between 0.83 and 6.7 $\sigma$.  
Using the fact that the Bjerrum length in water at 25~$^o$C is about 
7.1~\AA, the typical range of $\lambda_B$ in an aqueous solution would be
approximately 1 to 2 $\sigma$ if the monomer size $\sigma$ 
is taken to be roughly 4 to 8~\AA.

We employ standard molecular dynamics (MD) methods \cite{MD} for the simulation.
To carry out constant temperature dynamics, either Brownian dynamics 
\cite{MD-BD} or stochastic collision \cite{MD-SC} can be used.
We have found stochastic collision to be more efficient, because it generally 
generates a faster relaxation time for the polymer configurations  
\cite{drovetsky}.  In our simulations, we assign new velocities to all the
particles from a Maxwell-Boltzmann distribution every 100 to 1000 MD time 
steps.

The dynamics of the system is performed using the Verlet 
leap-frog/central difference algorithm at a time step of
$0.015 \tau$ where $\tau=\sigma\sqrt{m/48\epsilon}$.  
To ensure full equilibration, 
the simulations were run long enough such that the chains move at least six 
times their contour lengths.  This required anywhere from 250,000
to 4,000,000 time steps for most systems studied.

We have studied 
multi-chain systems at three different monomer number densities: 
$\rho \sigma^3$ = $10^{-3}$, $10^{-4}$ and $10^{-5}$, 
with various chain lengths $N$ = 16, 24, 32, and 48.  
The system sizes are chosen
so that the contour length of each chain does not exceed half of
the box length.  This translates to 12 chains for $N$=16, 10 for 
$N$=24, 6 or 9 chains for $N$=32, and 6 for $N=48$.  These systems
are all well below the overlap density.  
We have also performed extensive 
simulations with 27 chains at the above densities 
with $N=10$.

\section{Results and Discussions}\label{sect3}

\subsection{Counterion Condensation}\label{sect3a}

Counterion condensation is difficult to quantify directly.  Counting 
the number of counterions within a certain distance from the polyelectrolyte
is a possible though not very precise measure since the macroion can have
influence on a counterion far away from the chain due to the long-range
nature of the coulomb potential.
As an alternative, the osmotic pressure has often been used, both
experimentally and theoretically, as an indirect measure of the degree 
of counterion condensation.  
To truly determine the degree of counterion condensation though, 
it is necessary to study the correlation between the motions of the
counterions and that of the macroion.  This is best done by examining
the velocity correlation between the counterions and the 
chain.  

In lieu of measuring the velocity cross-correlation, 
there is a simpler alternative for quantifying counterion condensation.  
We can measure the diffusion coefficient of the
counterions instead.  If the chains are massive, they diffuse slowly.  
Therefore, if the motions of the condensed counterions are correlated 
to the chains, they would exhibit a substantially reduced diffusion 
coefficient too.  Measuring the diffusion coefficient of the counterions 
$D_C$ is in principle easy, at least theoretically, 
and this provides an accurate 
estimate of the degree of counterion condensation.  
In the limit where the uncondensed counterions
are essentially ideal, i.e. assuming that they do not interact with
each other strongly, $D_C$ should be simply proportional to the number of 
uncondensed counterions.
The counterion diffusion coefficient $D_C$ can also be related to the 
mobility $\mu_C$ of the 
counterions via the Einstein relationship.

\begin{figure}
\epsfxsize=0.85\columnwidth
\centerline{\epsffile{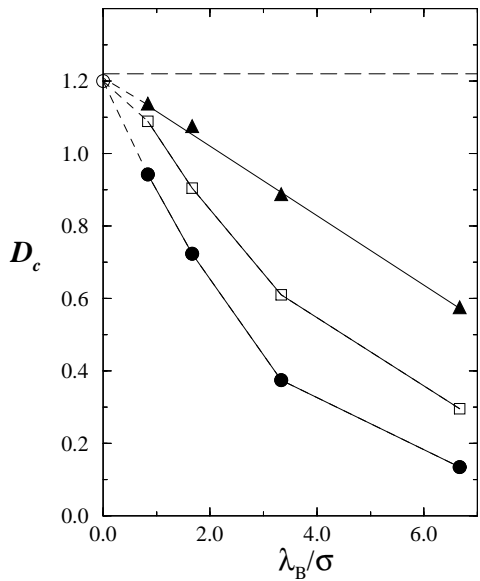}}
\caption[]{\label{fig-dc}
Self-diffusion coefficient $D_C$ of counterions at three different 
concentrations
$\rho\sigma^3$ = $10^{-3}$ (closed circles), $10^{-4}$ (open squares)
and $10^{-5}$ (closed triangles) as a function of Bjerrum length $\lambda_B$.
For an estimate of the intrinsic self-diffusion coefficient in the absence
of any counterion condensation, the open circle indicates $D_C$ obtained from a
simulation with the coulombic interactions turned off and
the dashed line indicates $D_C$ from simulations with counterions only but
no polyelectrolyte chains.
}
\end{figure}

The counterion diffusion coefficients $D_C$ are shown in Fig.~\ref{fig-dc}
for various Bjerrum lengths at the three different concentrations studied,
closed triangles for $\rho \sigma^3$ = $10^{-5}$, 
open squares for $10^{-4}$ and closed circles for $10^{-3}$.
In order to estimate the intrinsic mobility of the counterions in the
absence of counterion condensation, 
we have also run the simulations at these three concentrations for various
$\lambda_B$ {\em without} the macroions.
The dashed line in Fig.~\ref{fig-dc} indicates this intrinsic diffusion 
coefficient $D_C^0$, which was found to be independent of $\lambda_B$ for this
concentration regime and roughly independent of $\rho$ also.
The open circle at $\lambda_B$ =0 indicates the diffusion coefficient 
obtained for a system with the coulombic
interactions turned off, providing another 
independent estimate for the intrinsic 
diffusion coefficient $D_C^0$ in the absence of counterion condensation.

As expected, $D_C$ decreases with increasing $\lambda_B$, 
signaling an increasing counterion condensation.  
For the lowest concentration studied $\rho\sigma^3$ = $10^{-5}$ (triangles in 
Fig.~\ref{fig-dc}), the dependence is 
apparently linear, and this linearity appears to extend throughout the
entire range of $\lambda_B$ studied.  
For still larger values of $\lambda_B$ which we have not studied, 
this linearity is expected to eventually give way to a faster
than linear decrease, 
since $D_C$ obviously cannot be less than zero.
For the two higher concentrations, the dependence deviates from 
linearity sooner, but the basic features are the same.  

There is no sharp break in the
$D_C$ dependence on $\lambda_B$.  This suggests a behavior qualitatively
different from that predicted by Manning theory.
Manning theory 
suggests that there is some critical value of $\lambda_B = b$ below which 
no condensation occurs.  For the model used here, $b$, the mean distance
between charges on the chain, is 1.  According to Manning theory, the
onset of counterion condensation should then occur at $\lambda_B$ = 1.  However,
we observe in Fig.~\ref{fig-dc} that for all concentrations,
the dependence of $D_C$ on $\lambda_B$ does not exhibit any break
at $\lambda_B$ near 1 or at any other value.

\begin{figure}
\epsfxsize=0.85\columnwidth
\centerline{\epsffile{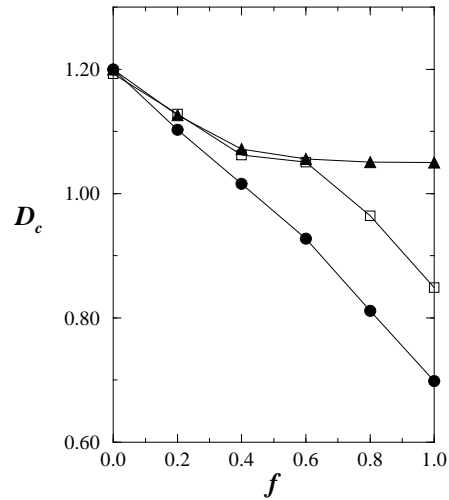}}
\caption[]{\label{fig-dc-f}
Self-diffusion coefficient $D_C$ of counterions in solutions containing
partially-charged polyelectrolyte chains with charge fraction $f$.
Three different concentrations are shown:
$\rho\sigma^3$ = $10^{-3}$ (closed circles), $10^{-4}$ (open squares)
and $10^{-5}$ (closed triangles) and $\lambda_B$ = 1.7.
}
\end{figure}

This departure from Manning theory is likely a result of the flexibility
in the chain conformation, which in turn is strongly dependent on the
degree of counterion condensation.  As we will show in Sect.~\ref{sect3b},
the structure of flexible polyelectrolytes are not always extended, and
this affects the validity of the arguments behind Manning theory, 
which was originally formulated for long rods.

\begin{figure}
\epsfxsize=0.85\columnwidth
\centerline{\epsffile{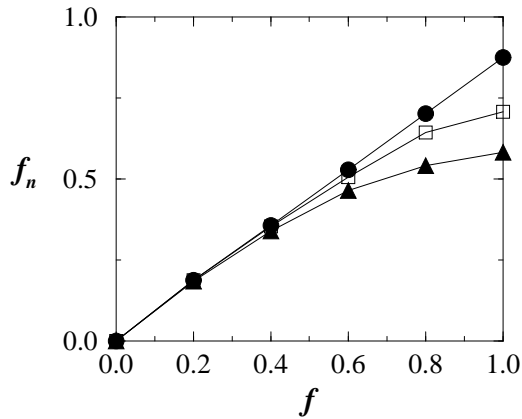}}
\caption[]{\label{fig-fn-f}
The renormalized charge fraction $f_n$ on each chain as a function
of the charge fraction on the bare chain extracted from counterion
diffusion coefficient data from Fig.~\ref{fig-dc-f}.
Three different concentrations are shown:
$\rho\sigma^3$ = $10^{-3}$ (closed circles), $10^{-4}$ (open squares)
and $10^{-5}$ (closed triangles) and $\lambda_B$ = 1.7.
The data do not exhibit the saturation behavior predicted by Manning theory.
}
\end{figure}

Using the simulations, we can examine other qualitative aspects of 
Manning theory.  
For example, one can study the condensation of counterions onto 
partially-charged polyelectrolyte chains.
The results presented in Fig.~\ref{fig-dc} are for
fully-charged chains, i.e. every monomer on the backbone is charged.
We have also examined partically-charged chains. 
By charging only certain fractions of the
monomers, we simulated solutions of chains with different charge fractions $f$.
The condensation of counterions was again quantified by measuring the 
counterion diffusion coefficients $D_c$.  
According to Manning theory, 
counterion condensation will continue until the renormalized charge
fraction $f_n$ (charge of the backbone minus the charge of the 
condensed counterions divided by the number of monomers) on the chain 
is such that the average charge
separation is equal to the Bjerrum length $\lambda_B$.  Consequently, 
Manning theory predicts that the renormalized charge fraction $f_n$
as a function of the charge fraction of the bare chain $f$ would exhibit a 
saturation behavior.  Figure~\ref{fig-dc-f} shows how $D_c$ depends on
$f$ for three different concentrations at $\lambda_B$ = 1.7.
From these data we can determine the number of condensed counterions and 
compute the renormalized charge fraction $f_n$.  Figure~\ref{fig-fn-f}
shows $f_n$ extracted from results in Fig.~\ref{fig-dc-f} plotted as a 
function of $f$.  For this set of parameters, the renormalized charge 
fraction $f_n$ should saturate at 0.6 according to Manning theory.  
But the data show no such saturation.  Instead, the renormalized charge 
fraction $f_n$ seems to increase monotonically for all concentrations
studied.

Our observation that there is no saturation in the renormalized charge
fraction is consistent with the analytical 
theory of Liu et al. \cite{becky} for finite-length polyelectrolyte rods 
in a poor solvent.  However, a recent small-angle X-ray and 
neutron scattering experiment \cite{williams} seems to suggest
that saturation does occur, in support of Manning theory.  
This experimental conclusion was based on the 
assumption that the peak observed in the scattering amplitude arose 
from a correlation tube of the size of the Debye-H\"uckel screening
length around each chain.  Since the screening length is thought to 
depend on the
concentration of free counterions, it was argued that the position of the
scattering peak could be used to determine the renormalized charge
fraction on the chains.  In Sect.~\ref{sect3c}, we will show that
this interpretation may be incorrect.  

\begin{figure}
\epsfxsize=0.85\columnwidth
\centerline{\epsffile{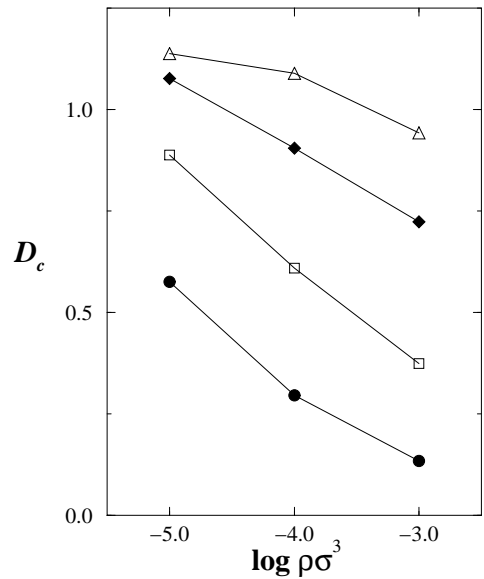}}
\caption[]{\label{fig-dc-a}
Data from Fig.~\ref{fig-dc}, replotted as a function of concentration
for different
Bjerrum lengths $\lambda_B$ = 0.83 (triangles), 1.7 (diamonds), 3.3 (squares)
and 6.7 (circles).
}
\end{figure}

\begin{figure}
\epsfxsize=0.85\columnwidth
\centerline{\epsffile{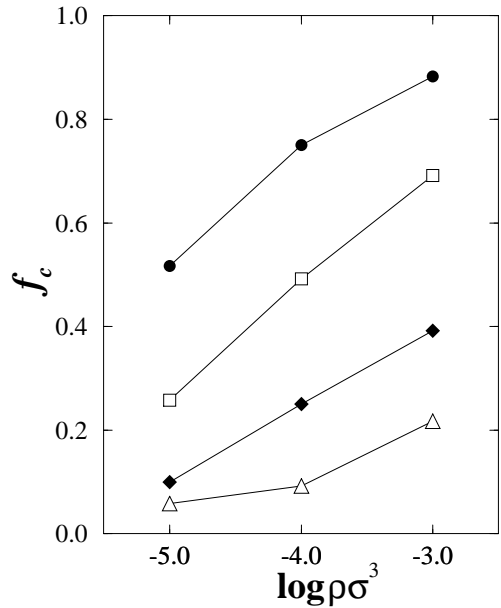}}
\caption[]{\label{fig-fc-c}
The fraction of condensed counterions extraced from data 
in Fig.~\ref{fig-dc-a} for different
Bjerrum lengths $\lambda_B$ = 0.83 (triangles), 1.7 (diamonds), 3.3 (squares)
and 6.7 (circles).
}
\end{figure}

The counterion diffusion coefficient in Fig.~\ref{fig-dc} are replotted in 
Fig.~\ref{fig-dc-a} as a function of concentration for the four different 
Bjerrum lengths studied.  At each $\lambda_B$, there is
a steady decrease in $D_C$ with increasing concentration, 
indicating that counterion condensation increases with increasing 
concentration.  Again, the dependence is smooth, without any abrupt changes
or any signs of a critical transition point.  Our results are 
qualitatively consistent
with the experimental measurements of counterion self-diffusion coefficients 
using pulsed field gradient NMR techniques in the low concentration region
\cite{schipper}.
From these data, we can directly determine the fraction of 
counterions condensed onto the chains.
This is shown in Fig.~\ref{fig-fc-c} as a function of concentration 
for the four different $\lambda_B$ studied.  
Clearly, the number of counterions condensed is not only a function of
the Bjerrum length, but also a strong function of concentration.

\begin{figure}
\epsfxsize=0.85\columnwidth
\centerline{\epsffile{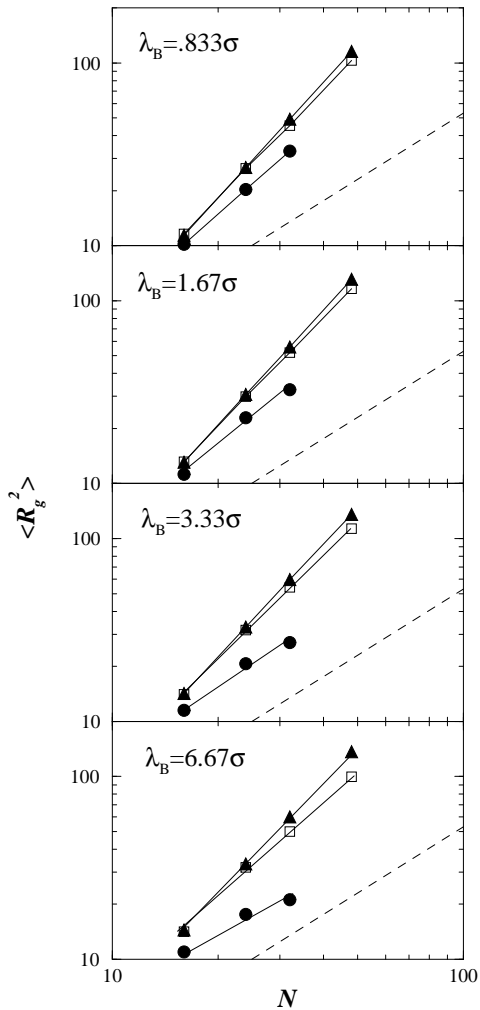}}
\caption[]{\label{fig-rg2}
Log-log plot of the average squared radii of gyration as a function of chain
length $N$ for different Bjerrum lengths and different concentrations 
$\rho\sigma^3$ = $10^{-3}$ (circles), $10^{-4}$ (squares) and $10^{-5}$
(triangles).  The dashed lines indicate results from simulations with
the coulombic interactions turned off.
}
\end{figure}

\subsection{Single-Chain Conformations and Intrachain Attractions}
\label{sect3b}

We have studied the single-chain structure at various concentrations
and Bjerrum lengths for a number of different chain lengths $N$.
The equilibrium averaged squared radii of gyration $\langle R_g^2\rangle$
are shown in Fig.~\ref{fig-rg2} for four different Bjerrum lengths.  
Data for the three different concentrations are displayed, with triangles 
for $\rho \sigma^3$ = $10^{-5}$, squares for $10^{-4}$ and circles for
$10^{-3}$.  For all three concentrations, the data fall on straight lines
on a log-log plot for all $\lambda_B$ studied, 
indicating that the chain lengths from the shortest one with $N=16$ 
on are all in the scaling limit.  
For comparison, we have also shown the squared radii of gyration 
for neutral chains at infinite dilution as dashed lines in Fig.~\ref{fig-rg2}.
For all lengths studied, the charged chains are substantially larger
than the neutrals.

From the slopes of the lines in Fig.~\ref{fig-rg2} we can extract the exponent
$\nu$ in the scaling relationship $\langle R_g^2 \rangle \sim N^{2\nu}$, 
and their values are shown in Table~\ref{tab1}.  For the lowest concentration 
$\rho \sigma^3$ = $10^{-5}$, the chains are fully extended with 
$\nu \approx 1$ for all $\lambda_B$.  For the intermediate concentration 
$\rho \sigma^3$ = $10^{-4}$,
the chains are still fairly extended, with a slight 
variation in $\nu$ going toward larger $\lambda_B$.  

\begin{figure}
\epsfxsize=0.85\columnwidth
\centerline{\epsffile{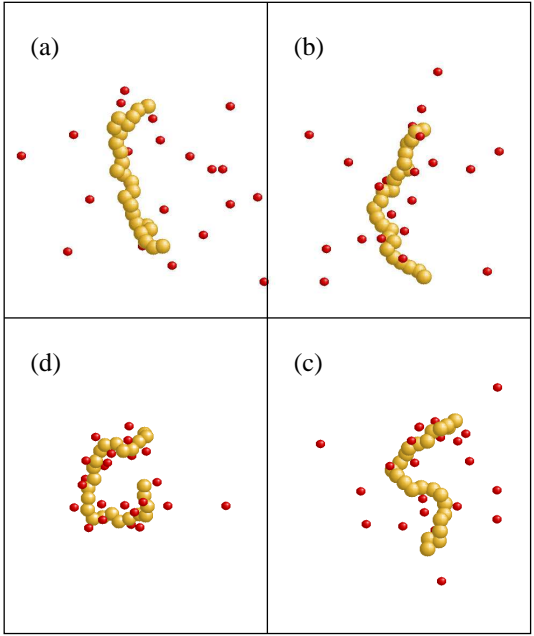}}
\caption[]{\label{fig-chains}
Typical chain conformations for concentration $\rho\sigma^3$ = $10^{-3}$
at four different Bjerrum lengths, $\lambda_B$: (a) 0.83$\sigma$, (b) 1.7$\sigma
$,
(c) 3.3$\sigma$ and (d) 6.7$\sigma$.  Light-colored spheres represent
monomer units on the polyelectrolyte chains and dark-colored spheres the
counterions.  (Differences in the size of the spheres are for display
purpose only.  The counterion and monomers have identical effective size
in the simulations.)  Notice the variations in the degree of counterion
condensation as a function of $\lambda_B$.
}
\end{figure}

For the highest concentration $\rho \sigma^3$ = $10^{-3}$, however,
the variation in the
exponent $\nu$ is quite substantial.  For the smallest $\lambda_B$ = 0.83, 
$\nu \approx$ 0.85, indicating a semi-extended chain structure.  But at the
largest $\lambda_B$ = 6.7, $\nu \approx$ 0.48, indicating a globular 
structure with a scaling exponent 
that is actually {\em smaller than a gaussian chain}.  In between these
two extremes, there is a consistent decrease in $\nu$, suggesting that
the size of the chain decreases steadily toward more globular structures.  
Figure~\ref{fig-chains} shows examples of chain conformations at $\rho\sigma^3$
= $10^{-3}$ for different Bjerrum lengths, illustrating the progression
from a semi-extended structure to a globular structure in accordance with 
the variation in $\nu$.
Though the scaling indicates an exponent smaller than gaussian behavior
for 
\vbox{
\begin{table}[h]
\caption[]{Scaling exponents for $\langle R_g^2 \rangle$
at various Bjerrum lengths and concentrations.
\label{tab1}
}
\begin{tabular}{rrr}
\multicolumn{1}{c}{$\lambda_B$} & \multicolumn{1}{c}{$\rho$}
& \multicolumn{1}{c}{$\nu$} \\
\hline
0.833 & $10^{-3}$ & 0.84 \\
      & $10^{-4}$ & 0.98 \\
      & $10^{-5}$ & 1.0 \\
1.667 & $10^{-3}$ & 0.77 \\
      & $10^{-4}$ & 0.99 \\
      & $10^{-5}$ & 1.0 \\
3.333 & $10^{-3}$ & 0.62 \\
      & $10^{-4}$ & 0.94 \\
      & $10^{-5}$ & 1.0 \\
6.667 & $10^{-3}$ & 0.48 \\
      & $10^{-4}$ & 0.82 \\
      & $10^{-5}$ & 1.0 \\
\end{tabular}
\end{table}
}
large values of $\lambda_B$, 
it is important to remember that 
the absolute size of the charged chains are 
larger than the neutrals (dashed lines in Fig.~\ref{fig-rg2}).

Recent studies of effective attractive interactions between 
like-charged polyelectrolytes have 
considered only rigid rods \cite{gelbart,liu2} where 
the issue of intrachain attraction did not come up.  A natural extension 
of these studies is to ask 
whether an effective attractive interaction also exists between 
different parts on the same chain, and if it does, whether
its nature is identical to the interchain interactions.
The globular structures shown in Fig.~\ref{fig-chains} for large Bjerrum lengths
suggest the existence of attractive intrachain interactions.  This is 
corroborated by the smaller-than-gaussian scaling exponent $\nu$ observed.
Compared to the extended structures at small $\lambda_B$, 
the higher $\lambda_B$
globular structures are more contorted, with a larger number of bends.
The same behavior has also been observed in a previous
simulation of flexible polyelectrolytes \cite{stevens-kremer}. 
The explanation ascribed to it was that increasing counterion condensation
at larger $\lambda_B$ screened the intrachain repulsion in such a way that
the structure of the chains went from being fully extended to that of an
essentially neutral polymer.
But the scaling evidence from the present work goes beyond this assertion
and suggests the presence of an attractive interaction
causing the chains to attain a structure {\em even more compact than
that of a neutral polymer}.  This collapse of the chain agrees qualitatively
with the recent results of Winkler et al. \cite{winkler}

Having established the presence of an attractive intrachain interaction,
the next question is whether this attractive interaction has the same
origin as the interchain attraction observed in polyelectrolyte rods.
This is a difficult question to answer.  Interchain attraction can 
arise from the charge fluctuations along the backbone of the 
rods as condensed counterions come on and off the rods.  But this is
not the only mechanism possible.  Ray and Manning recently suggested that two
like-charged rods can share counterions in such a way to form what is
analogous to a covalent bond \cite{covalent}.  The effective distance
scale of this covalent-like interaction is predicted to be of the order
of the Debye-H\"uckel screening length, which is much longer 
than the intrachain distance here.  Therefore, the first mechanism is 
more likely to be correct for the attractive intrachain interactions.
However, more work is necessary to fully elucidate the origin of the
intrachain attraction.

The attractive interactions are evident only for solutions with
very large Bjerrum lengths.  For smaller $\lambda_B$, the chains are apparently 
well extended.  For these cases, the intrachain attractions, if
present, are not strong enough to overcome the coulombic repulsion between
different parts of the same chain.

\subsection{Interchain Interactions}
\label{sect3c}

The most straightforward way to determine whether interchains attraction 
exists is to calculate the free energy as a function of the separation
between two chains.  Theoretically, this could be done by an umbrella
sampling procedure \cite{umbrella} or by computing the potential of
mean force of separating two chains \cite{mean-force}.
Unfortunately, due the long-range nature of the coulomb potential, 
both of these methods suffer from heavy statistical noise.  
Compounded by the need to deal with a finite polymer concentration
(i.e. multiple chains), these direct methods are not always very useful 
in practice.

An alternative is to monitor
the radial distribution function $g(r)$ of the centers-of-mass of the 
polyelectrolyte chains.  The logarithm of $g(r)$ is related to the
free energy of separating a pair of chains, so by monitoring $g(r)$
we can obtain information equivalent to those given by direct calculations
of the free energy.  From sufficiently 
long MD simulations carried out at equilibrium, an accurate 
determination of the radial distribution function is not too difficult.

The data we will show were obtained from very long simulations of $M=27$ 
chains with $N=10$ monomers at many different concentrations and Bjerrum 
lengths.  These chains are shorter than those used for obtaining the 
results shown in the last two subsections.  The motivation for this was 
to enhance the statistics in the $g(r)$ calculations by putting more
chains in the simulations, but reducing the size of each chain to keep
the computation at a manageable level.  We have also run calculations at
a number of different combinations of $M$ and larger $N$.  The results
are all qualitatively identical to those shown below for $M=27$ and $N=10$.

\begin{figure}
\epsfxsize=0.85\columnwidth
\centerline{\epsffile{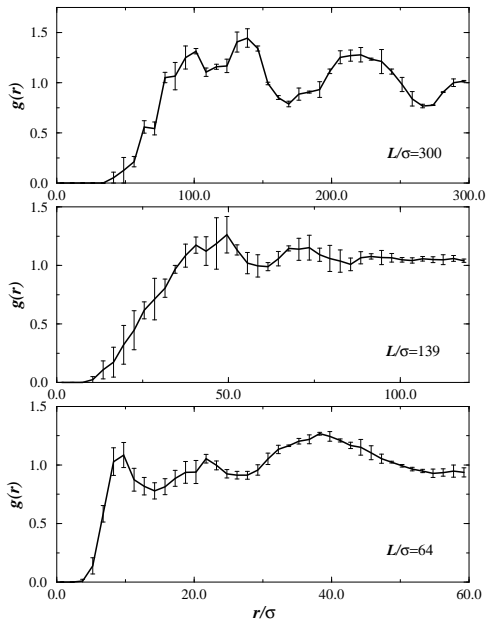}}
\caption[]{\label{fig-gr-0}
The polymer radial distribution function for the centers-of-mass of the
chains at Bjerrum length $\lambda_B$ = 0.83$\sigma$,
for three different concentrations 
$\rho\sigma^3$ = $10^{-5}$ ($L/\sigma$ = 300), $10^{-4}$ ($L/\sigma$ = 139)
and $10^{-3}$ ($L/\sigma$ = 64).
}
\end{figure}

Figure~\ref{fig-gr-0} shows the radial distribution function $g(r)$ as a
function of the center-of-mass separation between chains
for a very small Bjerrum length $\lambda_B = 0.83\sigma$.
The simulations were carried out for three different concentrations by
varying the size of the simulation box $L$: 
$\rho\sigma^3$ = $10^{-5}$ ($L/\sigma$ = 300), $10^{-4}$ ($L/\sigma$ = 139)
and $10^{-3}$ ($L/\sigma$ = 64).

Although we have used very long runs to extract the $g(r)$ functions, there
is still substantial statistical noise in the data.  We have indicated the
size of the statistical noise by error bars which represent one standard 
deviation from the mean.  Despite this, some of the minor peaks and
valleys in $g(r)$ may not be significant.  So we will rely on the gross
features of $g(r)$ to draw our conclusions.

For the lowest concentration ($L/\sigma$ = 300), $g(r)$ is fairly featureless
with two broad humps.
There is a correlation hole for distances smaller than roughly $100\sigma$,
indicating that chains cannot come closer together by distances much smaller
than $100\sigma$.
Beyond that, there is a fairly constant density.
Apparently, for this $\lambda_B$ at this concentration, 
the polyelectrolyte chains behave very much like particles with
an effective radius of approximate $100\sigma$.  Because there are 27
chains in the periodic box of length $L=300\sigma$, $100\sigma$ happens to
be the dimension of the average volume available for each chain.  So the
effective radius of $100\sigma$ deduced from the $g(r)$ appears to be
a result of a net repulsive interaction
pushing chains away from each other to fully fill the volume available.
It is important to point out here that the radius of gyration of the
chains are much shorter than the dimension of the correlation hole and
the concentrations studied here are all well below the overlap density.
Incidentally, the Debye-H\"uckel screening length for these conditions is also
roughly $100\sigma$.  Therefore, it is also possible that the correlation
hole in $g(r)$ may be related to the screening length instead.  That this
is actually not the case will be evident from data to be presented below.

Now turning to the intermediate concentration $\rho\sigma^3$ = $10^{-4}$ 
($L/\sigma$ =139), we observe similar behavior as for the lower concentration.
This time, the dimension of the correlation hole shrinks to about 
$50\sigma$, which again is about 1/3 of $L$.  Beyond this, $g(r)$ is again
rather featureless, indicating an essentially structureless solution of chains
with repulsive interactions.

\begin{figure}
\epsfxsize=0.85\columnwidth
\centerline{\epsffile{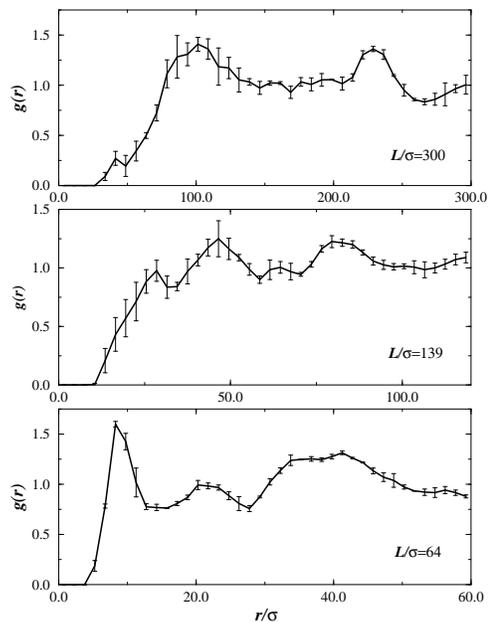}}
\caption[]{\label{fig-gr-1}
Same as Fig.~\ref{fig-gr-0} for $\lambda_B$ = 1.7$\sigma$.
}
\end{figure}

Next, we examine the highest concentration $\rho\sigma^3$ = $10^{-3}$ 
($L/\sigma$ = 64).  $g(r)$ shows a distinct peak in the short-distance
region.  Beyond that, $g(r)$ has the generally featureless appearance
similar to the last two concentrations.  Based on the observations
in the previous two cases, we expect the dimension of the correlation hole
to be at 1/3 the box size, namely around $20\sigma$.  
But the left edge of $g(r)$ now stands at approximately $r = 10\sigma$.  
These facts together indicate that the peak at small distance is a new 
feature not present in the previous cases,
and the left edge of $g(r)$ does not actually correspond to the
dimension of the correlation hole.  
The origin of this new feature will become clear below.

In Fig.~\ref{fig-gr-1}, we show the same data for a larger 
$\lambda_B$ = $1.7\sigma$.
For the lowest concentration, a similar $g(r)$ is observed here compared to 
the one at the smaller Bjerrum length in Fig.~\ref{fig-gr-0}.  
Notice that the dimension of the correlation hole has not changed, although 
the screening length is smaller for $\lambda_B$ = $1.7\sigma$ than it is for 
$0.83\sigma$ in Fig.~\ref{fig-gr-0} by a factor of $\sqrt{2}$.
So the size of the correlation hole is apparently unrelated to 
the Debye-H\"uckel screening length. 

The size of the correlation hole appears to be more closely 
related to the polymer concentration instead of the screening length.
Essentially the same conclusion can be inferred from the study of 
Stevens and Kremer \cite{stevens-kremer}, which showed that in the very low 
concentration regime, the position of the peak in the structure factor
scales with the polymer concentration to a power of approximately 1/3.  
In this very dilute regime, the peak in the structure factor is
entirely due to the correlation hole in $g(r)$; therefore, the size of
the correlation hole should scale with the polymer concentration to a
power of $-1/3$.  This is corroborated by our observations here 
that the correlation hole shrinks by a factor of roughly 2 when the density
decreases by a factor of 10.

The fact that the size of the correlation hole has no apparent relationship
to the screening length has important implications for the small-angle
X-ray and neutron scattering experiments \cite{williams}
referred to in Sect.~\ref{sect3a}.
Our results suggest that the arguments used in the experiments in arriving 
at the conclusion that the counterion condensation saturates at a certain
renormalized charge fraction is wrong, i.e. the size of the correlation hole 
extracted from the peak in the scattering intensity may have nothing to
do with the screening length at all.

For the intermediate concentration in Fig.~\ref{fig-gr-1}, 
there appears to be an extra feature
for distances smaller than the expected size of the correlation hole
($50\sigma$ as in Fig.~\ref{fig-gr-0}) just like in the highest concentration 
in Fig.~\ref{fig-gr-0}.  Given the size of the statistical
noise and the quality of the data, we are uncertain about the significance
of this feature.  But the same feature will become more prominent in
the data to be shown below.  So we skip to the highest concentration.

\begin{figure}
\epsfxsize=0.85\columnwidth
\centerline{\epsffile{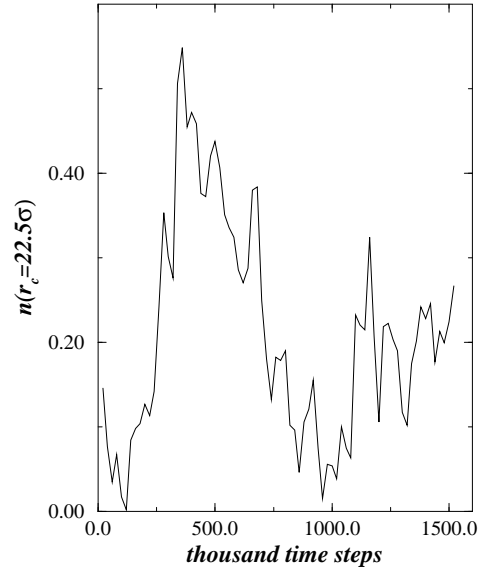}}
\caption[]{\label{fig-gr-trans}
Equilibrium fluctuations of the integrated number of chains under the first
peak in $g(r)$ up to a cutoff distance of $22.5\sigma$
from the data of Fig.~\ref{fig-gr-0} at $\rho\sigma^3$ = $10^{-4}$ as a
function of MD time steps.
}
\end{figure}

For the highest concentration in Fig.~\ref{fig-gr-1}, the feature 
observed previously for $\lambda_B$ = $0.83\sigma$
is now much more pronounced.  There is now a sharp peak at $r\approx 10\sigma$.
This peak is indicative of some kind of clustering of two or more chains
at short distances.  
By integrating under this peak, we can estimate the average number of
neighbors each chain has.  
This turns out to be
approximately 0.41.  This is strong evidence that there exists 
an attractive interaction between chains, causing them to cluster with 
each other.  Indeed, if we examine the configurations, we observe
a small number of chains pairing up.  These pairs appear to be
true equilibrium structures, because they would dissolve and reform
continually, even in cases where the short-distance peak in $g(r)$ cannot
be clearly resolved.  For example, 
the integrated peak height in the radial distribution
function up to $r=22.5\sigma$ is plotted in Fig.~\ref{fig-gr-trans} for 
$\rho\sigma^3$ = $10^{-4}$ and $\lambda_B$ = $0.83\sigma$, indicating that
the clustering is a transient but equilibrium phenomena.

\begin{figure}
\epsfxsize=0.85\columnwidth
\centerline{\epsffile{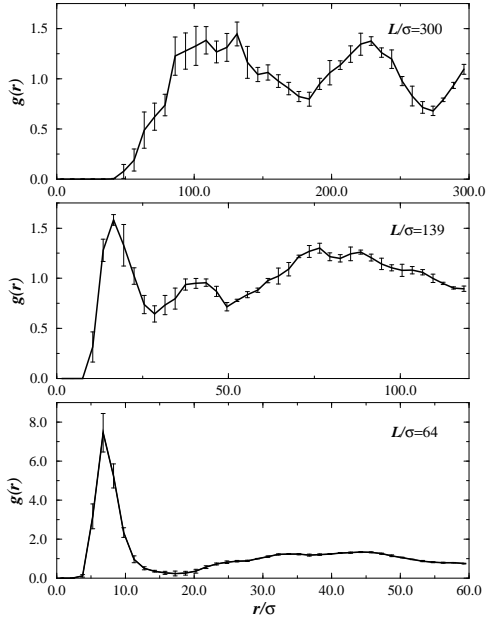}}
\caption[]{\label{fig-gr-3}
Same as Fig.~\ref{fig-gr-0} for $\lambda_B$ = 3.3$\sigma$.
}
\end{figure}

\begin{figure}
\epsfxsize=0.85\columnwidth
\centerline{\epsffile{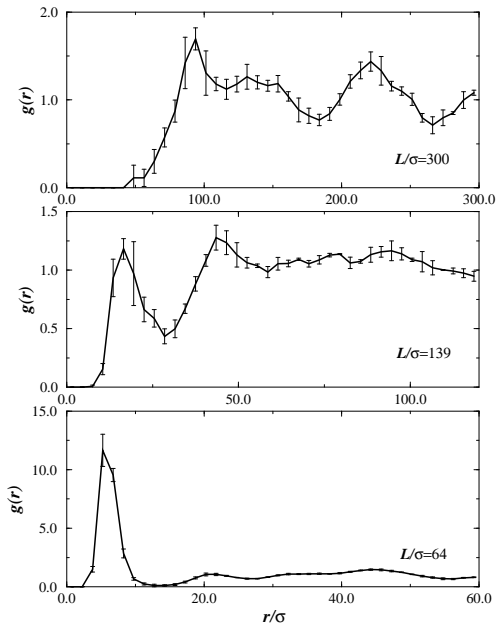}}
\caption[]{\label{fig-gr-6}
Same as Fig.~\ref{fig-gr-0} for $\lambda_B$ = 6.7$\sigma$.
}
\end{figure}

Figures~\ref{fig-gr-3} and \ref{fig-gr-6} show the same data for two larger
Bjerrum lengths $\lambda_B$ = $3.3\sigma$ and $6.7\sigma$.  For the two
highest concentrations, the sharp peak at short distance continues to grow
in intensity.  The feature that was barely visible for the intermediate
concentration in Fig.~\ref{fig-gr-1} is now fully developed.  The rise in the
short-distance peak indicates that more chains are clustering 
as either the concentration or the Bjerrum length or both
increase, implying that the attractive interactions grow in strength
at the same time.  The rising peak in $g(r)$ suggests that more chains
are participating in these clusters.
Again, by integrating under the short-distance peak in $g(r)$, we 
\vbox{
\begin{table}[h]
\caption[]{Average number of neighbors around each chain obtained from
integrating under the first peak in the radial distribution function
up to cutoff distance $r_c$.
\label{tab3}
}
\begin{tabular}{rrrr}
\multicolumn{1}{c}{$\rho$} & \multicolumn{1}{c}{$r_c/\sigma$}
& \multicolumn{1}{c}{$\lambda_B$} & \multicolumn{1}{c}{$n(r)$} \\
\hline
$10^{-3}$ & 9.75 & 0.867 & 0.29 \\
          & &  1.667 & 0.41 \\
          & &  3.333 & 1.69 \\
          & &  6.667 & 1.88 \\
                           \\
$10^{-4}$ & 25.5 &  0.867 & 0.32 \\
          & &  1.667 & 0.39 \\
          & &  3.333 & 0.74 \\
          & &  6.667 & 0.58 \\
\end{tabular}
\end{table}
}
can 
estimate the average number of neighbors each chain has, and the results
are tabulated in Table~\ref{tab3} for the two 
highest concentrations.
For instance, the peak in $g(r)$ for $\lambda_B = 6.7$ and concentration
$\rho\sigma^3 = 10^{-3}$ 
indicates that each chain has approximately 1.8 neighbors
on the average within a distance of $10\sigma$ around it.

\begin{figure}
\epsfxsize=0.5\columnwidth
\centerline{\epsffile{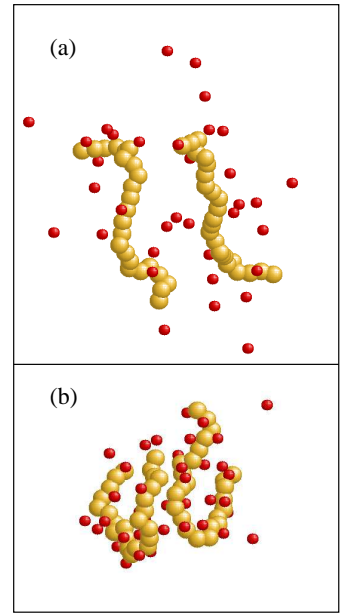}}
\caption[]{\label{fig-pairs}
Examples of chains pairing at two different Bjerrum lengths
$\lambda_B$ = (a) 3.3$\sigma$ and (b) 6.7$\sigma$ at the highest
concentration $\rho\sigma^3$ = $10^{-3}$.  Notice the high degree of
interchain correlations.
}
\end{figure}

Notice that in spite of the sharp peak observed at short distance, there are 
no other distinct secondary peaks.  If a large number of chains come together
to form a large-scale structure, there would have been second and third neighbor
peaks evident in $g(r)$.  The absence of these peaks suggests that most of the
intensity under the sharp peak at short distance is in fact due to small
clusters, in agreement with the data in Table~\ref{tab3}.
Indeed, a visual examination of the configurations indicates that most of the
chains are forming pairs or triplets.  Examples of pairs are shown in 
Fig.~\ref{fig-pairs}.
The precise reason for the absence of larger clusters 
is not clear at this point.  There are several possibilities:
(a) Some special feature of the attractive interchain interaction causes 
chains to form small clusters but precludes the creation of larger clusters;
(b) The number of chains used in the simulations is too small to 
accurately reflect the presence of large-scale structures; or 
(c) The lengths of our simulations are not long enough for the true equilibrium 
large-scale structures to emerge.  Larger simulations are needed to determine
the actual reason for the apparent absence of large clusters.

Also interesting is the $\lambda_B$ dependence of the position of the 
small-distance peak.  This is most clearly seen for the highest
concentration.  Starting from $\lambda_B$ = $0.83\sigma$, the peak position
shifts consistently to smaller values, from about $10\sigma$ to $6\sigma$
at $\lambda_B$ = $6.7\sigma$.  A similar trend can be seen in data for the
intermediate concentration.

The data in the radial distribution function demonstrate unequivocally 
that attractive interchain interactions do exist in solutions of 
like-charged flexible polyelectrolytes.  The strength of this attraction
also grows with polyelectrolyte concentration and the Bjerrum length.
These attractive interactions lead to clustering of chains, which
should be observable in neutron and light scattering experiments.  
Indeed, it has been suggested that the broad peak detected in recent 
neutron and light scattering 
experiments \cite{amis} at small wavevectors may actually be related to the
formation of clusters or domains \cite{ermi}.

To verify the connection between the short-distance peak observed
in the polymer center-of-mass radial distribution function and the broad 
low-wavevector peak detected in the scattering intensity, 
we can compute the structure factor $S(q)$ from the simulation data 
and compare them to the experiments.  Unfortunately, due to the small
size of the simulation box used, the number of allowed wavevectors 
dictated by the periodic boundary condition we have used is extremely
sparse at the low-wavevector region.  Instead of Fourier transforming
the monomer density directly to get $S(q)$, we first computed the 
monomer-monomer radial distribution function $g_m(r)$ 
and then $S(q)$ from it using the formula $S(q) = 1 + \rho_m h(q)$, 
where $h(q)$ is the Fourier transform of $h(r) = g_m(r) - 1$ and 
$\rho_m$ is the
monomer density.  We have compared the $S(q)$ computed this way with a
direct Fourier transform of the density, and found the qualitative features 
to be identical.

\begin{figure}
\epsfxsize=0.95\columnwidth
\centerline{\epsffile{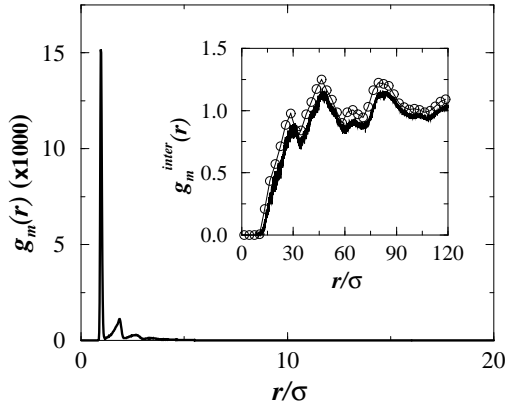}}
\caption[]{\label{fig-gm}
Total monomer-monomer radial distribution function $g_m(r)$ for
$\lambda_B = 1.7$ and $\rho\sigma^3$ = $10^{-3}$.
Inset shows the interchain contribution to $g_m(r)$ (solid line)
and the polymer center-of-mass radial distribution function (circles)
taken from Fig.~\ref{fig-gr-1}.
}
\end{figure}

\begin{figure}
\epsfxsize=0.70\columnwidth
\centerline{\epsffile{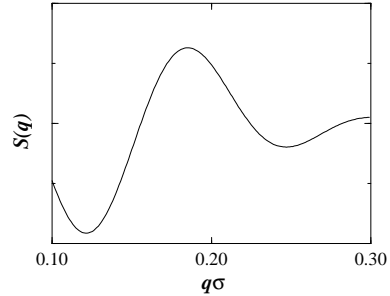}}
\caption[]{\label{fig-sq}
Structure factor derived from the interchain monomer radial distribution
function in Fig.~\ref{fig-gm} showing broad peak similar to those observed
in scattering experiments.
}
\end{figure}

Using the monomer radial distribution function offers another advantage.  
It allows us to separate the inter- and intra-chain contributions to the total 
scattering intensity.  Figure~\ref{fig-gm} shows the total monomer $g_m(r)$
for $\lambda_B = 1.7$ and $\rho\sigma^3 = 10^{-3}$.  Clearly, the total
(intra-chain plus inter-chain) $g_m(r)$ is dominated by the intrachain 
correlations at short distance.  The inset shows only the inter-chain 
contribution to $g_m(r)$ at a much expanded scale as the solid line.
Also in the inset, superimposed on the monomer $g_m(r)$ is the polymer
center-of-mass $g(r)$ for the same $\lambda_B$ and concentration taken from 
Fig.~\ref{fig-gr-1}.
Clearly, the interchain contribution to $g_m(r)$ closely resembles the 
polymer $g(r)$ we discussed earlier.  From the interchain contribution to
$g_m(r)$, we can then compute the part of the structure factor $S(q)$ that
arises solely from the interchain correlation, and this is shown in 
Fig.~\ref{fig-sq}.  The $S(q)$ shows a broad peak with a maximum at 
$q_{\rm max} \approx 0.18\sigma^{-1}$.  By truncating the Fourier
integral at different upper limits, we can verify directly that the
peak in $S(q)$ is essentially due to the first and second peaks in $g_m(r)$.
We found similar results for all other concentrations and Bjerrum lengths.

Although we have correlated the peak in the structure factor to the
attractive short-distance peak in the polymer $g(r)$, the accuracy of
our data precludes a reliable determination of how 
$q_{\rm max}$ scales with the polymer concentration.  
The reason for this is that although we have seemingly circumvented the 
problem of the sparseness of the allowed wavevectors by analyzing the monomer 
radial distribution function, the periodicity in principle does not
permit us to extract real independent information about $S(\vec{q})$ for 
wavevectors $q$ which do not correspond to reciprocal vectors of the cubic
boundary condition we employ.  If the actual peak lies between two
reciprocal vectors, we would not be able to accurately measure its position.
Given the size of the simulation box we have used for these calculations, 
the reciprocal vectors are certainly too sparse.  A much larger simulation
would be required for a direct comparison with experimentally obtained
scattering intensities.

Finally, 
we can compare the general characteristics of the attractive interactions
seen in our simulations of flexible polyelectrolytes with those of the
rigid rods observed in previous theoretical work \cite{gelbart,liu2}.  
First, the
attractive interactions between the flexibles are strongly concentration
dependent whereas for the rods they exist even for infinite dilution.
This is not really very surprising, given the fact that the attractive
interactions are results of charge fluctuations due to the condensed
counterions.  For infinite rods, condensation could occur even at
zero concentration, but for flexibles it could not.  
The degree of counterion condensation for flexibles 
increases with concentration, and so should the strength of the attractive
interactions. 
Secondly, the distance scale of the attractive interactions seems to be
very different in the flexibles compared to the rods.  For rods, 
previous calculations show that the attractive interactions are very
short-ranged, of the order of 10 or 20 \AA.  For the flexibles, if we
assume a monomer size of roughly 4 \AA, the typical position of the peaks 
observed in $g(r)$ from Figs.~\ref{fig-gr-0} to \ref{fig-gr-6} would be anywhere
from 30 to 100 \AA, depending on the concentration and the Bjerrum length.
On the other hand, these distances scales are very much in accord with 
those suggested by neutron and light scattering experiments 
\cite{amis,ermi,williams}.

\section{conclusions}

We have carried out detailed molecular dynamics simulations to study 
inter- and intra-chain interactions in solutions of flexible polyelectrolytes
at nonzero concentrations with no added salt.  
The simulations were performed with explicit 
counterions and unscreened coulombic interactions.  Specifically, we 
investigated questions related to the existence of effective attractive 
interactions as well as the nature of counterion condensation at finite
concentration.  

From the counterion self-diffusion coefficient, we 
were able to determine how counterion condensation depends on both the 
Bjerrum length and the polymer concentration.  The data revealed a
condensation mechanism that is distinctly different from that predicted
from Manning theory.  Specifically, we observed no saturation in the
renormalized charge fraction on the chains which appeared to be a smooth
function of the Bjerrum length as well as the polymer concentration.

From the scaling of the mean squared radii of gyration with the chain
length, we extracted the scaling exponent under a number of different
conditions with various Bjerrum lengths and concentrations.  For large
Bjerrum lengths and high polymer concentrations, a scaling exponent 
smaller than a gaussian chain was observed, indicating that the chains
had a more compact structure than ideal chains.  In conjunction with a
visual examination of the chain conformations, we concluded that 
the compact structures are results of attractive interactions between
different parts of the same chain.

To ascertain the presence of attractive interchain interactions, we have 
studied in great detail the radial distribution functions of the 
centers-of-mass of the chains in solution.  We found a peak at small
distances in the radial distribution function for high Bjerrum lengths
and/or high concentrations.  From the lack of secondary peaks in the
radial distribution function and a careful examination of the chain 
configurations, we concluded that this peak is a result of small domains
formed by chains clustering in the solution.  
This behavior is indicative of the presence of strong 
interchain attractions.  From the structure factor, we have directly verified
the connection of the small-distance peak in the polymer radial distribution 
with the broad peak detected at small wavevectors in the experimental 
small-angle X-ray and light scattering intensities.

\begin{acknowledgments}

Throughout this work, we benefited from numerous enlightening discussions
with Andrea Liu and Eric Amis.
This research has been supported in part by the National Science Foundation
under grants CHE-9528121 and CHE-9257094.
CHM is a NSF Young Investigator, 
a Camille and Henry Dreyfus Foundation Camille Teacher-Scholar and a
Alfred P. Sloan Foundation Fellow.
Computational resources have been
provided by the IBM Corporation under the SUR Program at USC.

\end{acknowledgments}

\end{document}